\documentstyle[12pt,epsfig]{article}

\newcommand{\onefig}[2]{
\begin{figure}[h]
\epsfig{ file=#1 ,
         height=8cm ,
         width=12cm } 
\end{figure}
}

\begin{document}
$\ $
\vskip 1.5 truecm

\centerline{\bf Bubble fluctuations in $\Omega<1$ inflation}

\vskip .5 truecm
\centerline{Jaume Garriga}
\smallskip
\centerline{\it IFAE, edifici C, Universitat Aut\`onoma de Barcelona}
\centerline{\it E-08193 Bellaterra, Spain}

\begin{abstract}

In the context of the open inflationary universe, we calculate the amplitude 
of quantum fluctuations which deform the bubble shape. 
These give 
rise to scalar field fluctuations in the 
open Friedman-Robertson-Walker 
universe which is contained 
inside the bubble. One can transform to a new
gauge in which matter looks perfectly smooth, and then
the perturbations behave as tensor modes (gravitational waves
of very long wavelength). For $(1-\Omega)<<1$, where $\Omega$ is
the density parameter, the microwave temperature anisotropies produced 
by these modes are of order $\delta T/T\sim H(R_0\mu l)^{-1/2}
(1-\Omega)^{l/2}$. Here,
$H$ is the expansion
rate during inflation, $R_0$ is the intrinsic radius of the bubble
at the time of nucleation, $\mu$ is the bubble wall tension and $l$ 
labels the different multipoles ($l>1$). The gravitational backreaction
of the bubble has been ignored.
In this approximation,
$G\mu R_0<<1$, and the new effect can be much larger 
than the one due to ordinary gravitational waves generated 
during inflation (unless, of course, $\Omega$ gets too close to
one, in which case the new effect disappears).

\end{abstract}

\section{Introduction}

The possibility of an open inflationary universe , in which the 
cosmological density parameter $\Omega$ is less than 1, has been
intensively studied in recent years \cite{martin,
linde,yama,lyth,tama,sasaki,selftama}. According to
this model, the universe is initially in a de Sitter phase, driven
by the potential energy of a scalar field trapped in a false 
vacuum $\sigma_f$ (see Fig. 1).
A bubble of 
the new vacuum $\sigma_t$ 
nucleates 
and its interior undergoes a second period of 
inflation. The homogeneity of our universe is then attributed
to the O(3,1) symmetry of the bubble: the interior of the light cone
from the nucleation event is isometric to an open 
Friedman-Robertson-Walker (FRW) universe \cite{coleman}.
The second period of inflation has to be suficiently long to generate the 
observed entropy, but if it is too long then $\Omega$ is driven 
exponentially close to 1. As
a result, to obtain
$\Omega<1$ the parameters of the scalar field potential have
to be fine tunned to some extent
\cite{martin,yama}.
Nevertheless, if observations 
ultimately determine that $\Omega$ is smaller than one, then 
open inflation may be regarded as a  `natural' scenario \cite{linde}.

The cosmic microwave background anisotropies produced by a nearly massless 
scalar field in open inflation were analyzed in \cite{yama}. It 
was shown that a `super-curvature' mode \cite{lyth}, which is not normalizable
on the open FRW space-like sections, would give a significant
contribution to the low multipoles if $\Omega<.1$ (see also \cite{
sasaki,ratra}).
A more complete study of cosmological perturbations in open inflation 
requires the quantization of fields in the presence of a bubble. Quantum field
theory in a bubble background was pioneered in \cite{tanmayalex}
and further developped in \cite{tama,sasaki,selftama}.
As noted in \cite{linde}, large contributions to microwave
perturbations may result from the quantum fluctuations of the bubble wall
itself \cite{jaumealex}. The purpose of this paper is to 
calculate the amplitude of such fluctuations and their effect on the microwave
background.

In Section 2 we briefly describe the bubble geometry. In 
Section 3 we calculate the amplitude of wall fluctuations for bubbles 
nucleated during inflation. This extends previous work for bubbles 
in flat space
\cite{jaumealex}(see also \cite{selftama}). In Section 4 we show that the 
effect of wall perturbations can be described in terms of long wavelength
tensor modes (analogous to
gravitational waves), and we evaluate their impact on the
microwave sky. Finally, in Section 5 we summarize our conclusions 
and compare them with recent related work \cite{linde,juan}.

\section{\bf Bubble geometry}

Before calculating the amplitude of bubble wall perturbations,
it will be useful to summarize, in this section, some of the features 
of the spacetime containing the bubble.
A conformal diagram
is given in
Fig. 2. The nucleation event is marked as N. The bubble 
wall is represented by the timelike hypersurface $w$ (solid line). In region I,
which is the interior of the 
light-cone from N, the line element is given by
\cite{coleman}
\begin{equation}
ds^2=-dt^2+a^2(t)d\Omega_{H^3}, \label{interior}
\end{equation}
where $d\Omega_{H^3}$ is the metric 
on the unit space-like hyperboloid
\begin{equation}
d\Omega_{H^3}=dr^2+\sinh^2r(d\theta^2+\sin^2\theta d\varphi^2).
\label{open}
\end{equation} 
Eq.(1) represents the 
geometry of an open FRW universe. This open universe inflates
up to the time when the scalar field reaches the value $\sigma=
\sigma_{rh}$ (see Fig. 1), and then reheats. After 
the usual radiation and matter dominated eras, it eventually 
becomes our observable universe. At all stages of expansion, the scale factor
obeys the Friedman equation
\begin{equation}
(1-\Omega)\dot a^2=1,
\label{friedman}
\end{equation}
where $\Omega$ is the ratio of the matter density $\rho_m$ 
to the critical density $\rho_c=(3/8\pi G)(\dot a/a)^2$.

The chart (\ref{interior}) covers only the interior of the light-cone from N.
One can cover the outside by analytically 
continuing the coordinates
$t$ and $r$ to the complex plane. By taking $t=i\tau$ and $r=\rho+i(\pi/2),$
where $\tau$ and $\rho$ are real, we have 
\cite{coleman}
\begin{equation}
ds^2=+d\tau^2+R^2(\tau)d\Omega_{dS}. \label{exterior}
\end{equation}
Here $R(\tau)=-i a(i\tau)$,  and
\begin{equation}
d\Omega_{dS}=-d\rho^2+\cosh^2\rho(d\theta^2+\sin^2\theta d\varphi^2)
\label{desitter}
\end{equation}
is the metric of a (2+1) dimensional de Sitter space of unit `Hubble length'. 
Note that $\tau$ is now a space-like coordinate, playing the role of a proper
radial distance from N. In this way,
the spacetime outside the light-cone from N is foliated 
into (2+1)-dimensional de Sitter leaves with constant $\tau$. In each one of 
these leaves $\rho$ plays the role of time.

The scalar field $\sigma$ obeys the field equation 
\begin{equation}
-\Box \sigma + {dV(\sigma)\over d\sigma}=0,
\label{fieldequation}
\end{equation}
where $\Box$ is the covariant d'Alembertian and $V(\sigma)$ 
is a potential of
the form depicted in Fig. 1.
The bubble configuration is a solution of (\ref{fieldequation}) of the
form 
$$\sigma=\sigma_0(\tau).$$
It has the shape of a kink that interpolates 
between the false vacuum $\sigma_f$ and the true one $\sigma_t$ 
\cite{coleman}. The locus where the field is at 
the top of the barrier, $\sigma(\tau_w)=\sigma_m$, can be identified 
with the 
trajectory of the domain wall (solid time-like 
line in Fig. 2). There, we have actually drawn an eternal bubble. 
A real bubble would nucleate 
at a given moment of time, say $\rho=0$, with intrinsic  radius 
\begin{equation}
 R_0=R(\tau_w). \label{r0}
 \end{equation}
The intrinsic radius 
would subsequently expand with $\rho$ as $R_0 \cosh \rho$. 
The scalar 
 field in region I can be found by analytically continuing from $\tau$ back
  to $t$ \cite{coleman}. 

Both the bubble configuration and the corresponding spacetime metric
enjoy an O(3,1) symmetry inherited from the spherical symmetry of the
instanton describing the tunneling \cite{coleman}. This is the group of 
isometries of the de Sitter leaves in region II and of the open 
hyperboloids in region I. For the scalar field, the symmetry simply 
means that $\sigma_0$ only depends on $\tau$ (or $t$).

 \section{\bf Scalar field fluctuations}

In this section we calculate the amplitude of small fluctuations 
for a bubble that nucleates during inflation. This extends previous 
results for bubbles in flat space \cite{jaumealex,tanmayalex,selftama}.
We shall work in the approximation in which the gravitational
 backreaction of the bubble can be ignored. In practice, 
 this means that up to the time of reheating, the energy density of matter
can be expressed as
 a large cosmological constant part $\Lambda/8\pi G$, plus a small 
 part $\delta \rho_m$ which contains the barrier feature of the potential 
 energy (see Fig. 1) plus the gradient and kinetic contributions of 
 the scalar field. Then we take the limit 
\begin{equation}
8 \pi G \delta \rho_m<<\Lambda. \label{app}
\end{equation} 
 The case when gravitational backreaction is included 
will be discussed elsewhere.

 In our approximation, the geometry during inflation 
 is that of de Sitter space.
 In region II the line element is given by (\ref{exterior}), with
 \begin{equation}
 R(\tau)=H^{-1}\sin(H\tau),\quad (0<\tau<\pi H^{-1}),
 \label{radius}
 \end{equation}
 where $H=(\Lambda/3)^{1/2}$ is the de Sitter Hubble rate. Note that $R$ 
 vanishes both at $\tau=0$ (the nucleation event N) and at $\tau=\pi H^{-1}$
 (the antipodal point A).
To study the wall fluctuations, we expand the scalar field as
\begin{equation}
\sigma(\tau, x^i)=\sigma_0(\tau)+\phi(\tau,x^i),
\label{small}
\end{equation}
where $x^i$ are the coordinates on the 
2+1-dimensional de Sitter leaves (\ref{desitter}). The small perturbation 
$\phi$ is promoted to a quantum operator $\hat \phi$, which is then expanded
into a sum over modes times the usual creation and anihilation 
operators
as $\hat \phi=\sum \phi_{klm} a_{klm}+ h.c.$ 
As noted in \cite{sasaki}, the spacelike surface $\rho=0$, connecting 
the nucleation event N with its antipodal A, is a good Cauchy surface
for the entire spacetime (see Fig. 2). 
Therefore we shall 
normalize our modes on that hypersurface.
Some of them, the so-called super-curvature modes \cite{yama,lyth}, 
may not be normalizable on the
open hyperboloids of the Friedman Robertson Walker chart  (\ref{open}), 
but this is just
because the hyperboloids are not good Cauchy surfaces. 
As we shall see, the perturbations of the bubble wall are super-curvature.

The equation of motion for small perturbations is
\begin{equation}
[-\Box+m^2(\sigma_0)]\phi_{klm}=0,
\label{perturbations}
\end{equation}
where $m^2(\sigma_0)={d^2V/d\sigma^2}|_{\sigma=\sigma_0(\tau)}$.
With the ansatz
\begin{equation}
\phi_{klm}=R^{-1}(\tau)F_k(\tau)Y_{klm}(x^i),
\label{mode}
\end{equation}
this separates into
\begin{equation}
[-^{(3)}\Box+k^2]Y_{klm}(x^i),
\label{living}
\end{equation}
and
\begin{equation}
-{d^2F_k\over d\eta^2}+R^2[m^2(\sigma_0)-2H^{2}]F_k=(k^2-1)F_k.
\label{schrodinger}
\end{equation}
Here $^{(3)}\Box$ stands for the covariant 
d'Alembertian on the 2+1 dimensional 
de Sitter leaves (\ref{desitter}), $k^2$ is a separation constant
and 
the conformal `radial' coordinate $\eta$ is  defined 
through the relation $R(\tau)d\eta\equiv d\tau$,
\begin{equation}
\cosh \eta\equiv {1 \over \sin(H\tau)}= {1\over H R(\tau)}.\label{eta}
\end{equation}

Equations 
(\ref{living}) and (\ref{schrodinger}) have a familiar interpretation 
\cite{tanmayalex,jaumealex}. The first one tells us that $Y_{k}$ behave as 
scalar fields of mass $k^2$ living in a 2+1-dimensional unit de Sitter space. 
The masses $k^2$ are determined as the eigenvalues of equation 
(\ref{schrodinger}),
which is simply a one dimensional Schrodinger equation with effective 
potential
\begin{equation}
U_{eff}=R^2[m^2(\sigma_0)-2H^2].\label{ueff}
\end{equation}
Note also that the modes 
$\phi_{klm}$ must obey the Klein-Gordon normalization condition
\begin{equation}
-i\int \phi_{klm} \stackrel{\leftrightarrow}{\partial_\mu}
\phi^*_{k'l'm'} d\Sigma^{\mu}=\delta_{kk'}
\delta_{ll'}\delta_{mm'},\label{kleingordon}
\end{equation}
where $\Sigma$ is the hypersurface $\rho=0$. 
If we choose the $Y_{klm}$ to be
Klein-Gordon normalized on the $2+1$ dimensional de-Sitter leaves, then
(\ref{kleingordon}) reduces to
\begin{equation}
\int_{-\infty}^{+\infty}F_kF_{k'}d\eta=\delta_{kk'},
\label{norm}
\end{equation}
which is the usual normalization condition for eigenfunctions of the 
Schrodinger equation.

The effective potential (\ref{ueff}) is schematically represented in Fig. 3.
The height of $U_{eff}$ at $\eta=0$ (which corresponds to $R(\tau)=H^{-1}$) 
is basically given by $(m_f^2H^2-2)$, where $m_f$ is the scalar field mass in 
the false vacuum. The narrow well on the left corresponds to the location 
of the bubble wall, where $m^2(\sigma)$ is negative and 
large in absolute value.
The equation of motion (\ref{fieldequation}) for $\sigma_0$ 
written in terms of the conformal coordinate $\eta$ is
$$
\sigma_0''+2{R'\over R}\sigma_0'-{dV(\sigma_0)\over d\sigma}=0,
$$
where primes denote derivatives with respect to $\eta$. 
Taking one more derivative with respect to $\eta$
it is straightforward to show that
\begin{equation}
F_{-3}\equiv N\sigma_0'(\eta) \label{theone}
\end{equation}
is a solution of (\ref{schrodinger}) with eigenvalue $k^2=-3$. This is 
analogous to what happens for bubbles in flat space 
\cite{jaumealex}. Note that $\sigma_0'=
R\dot \sigma_0$, where a dot indicates derivative with respect to the `radial' 
variable $\tau$. But in order for the instanton to be smooth, we must have
$\dot \sigma_0\to 0$ both at the nucleation event ($\eta\to -\infty$) and
at the antipodal point($\eta\to +\infty$) \cite{coleman}.
From (\ref{eta}), $R(\eta)$ also vanishes exponentially at 
$\eta\to\pm \infty$. 
Therefore
it is clear that the mode (\ref{theone}) is normalizable and that its 
eigenvalue $k^2=-3$ belongs to
the spectrum. In addition, since $\sigma_0$ is a monotonous function 
interpolating 
between true and false vacuum, the mode (\ref{theone}) has no nodes and is 
the eigenstate of lowest eigenvalue. Although all higher modes
will contribute to density perturbations and microwave temperature distorsions
\cite{sasaki},
for the remainder of this paper we shall focus on the 
lowest mode. This mode has a 
clear geometrical interpretation as deformations of the bubble shape,
which is the effect we are concentrating on.
To linear order we can write the perturbed field as
$$\sigma_0(\tau)+\phi_{-3}(\tau,x^i)\approx
\sigma_0(\tau+N Y_{-3lm}(x^i)).$$ 
Therefore, like in
the case of flat space \cite{jaumealex}, 
the perturbations associated with
$k^2=-3$ correspond to deformations that shift the position of 
the bubble wall in a $x^i$ dependent 
way, without altering the `radial' profile function $\sigma_0(\tau)$.

The normalization constant in (\ref{theone}) will 
eventually determine the magnitude of the effect.
In order to calculate it we use (\ref{norm}),
in the form
\begin{equation}
N^2\int R(\tau) \dot \sigma_0^2 d\tau =1.\label{norma}
\end{equation}
The integral can be numerically evaluated for any particular type
of bubble, but its meaning is best illustrated in the thin wall case. 
In this case
$\dot \sigma_0$ is very small except in a small region of size 
comparable to the width of the narrow well in Fig. 3, which is centered 
around the location of the wall, at $\tau=\tau_w$. The factor $R(\tau_w)$
can be pulled out of the integral and what remains is simply the wall
tension $\mu$. Therefore
$$
N^2={1 \over R_0 \mu},
$$
Here, as in (\ref{r0}),  $R_0=R(\tau_w)$ is the radius of the bubble at 
the moment of
nucleation. In the general case, the denominator in the right hand side 
is just shorthand for the integral in (\ref{norma}).

In the thin wall case, the explicit expression for the radius of the bubble 
at nucleation is given by (see e.g. \cite{jaumegros}),
\begin{equation}
R_0={3\mu\over (9\mu^2 H^2+\epsilon^2)^{1/2}},
\end{equation}
where $\epsilon$ is the jump in energy density between the true and the false
vacuum. As mentioned before, we have neglected the bubble's 
gravitational backreaction. For the approximation to be valid, we need on
one hand that $ G \epsilon<<\Lambda$ [see (\ref{app})]. On the other hand
we are neglecting the gravity of the wall. As is well known, the gravitational 
field of a domain wall is characterized by a Rindler-type horizon
distance \cite{alex}
\begin{equation}
l_w\equiv{1\over 8\pi G\mu}.\label{rindler}
\end{equation}
We need this distance to be much larger 
than the radius of the bubble at nucleation
\begin{equation} 
G\mu R_0<<1.\label{weak}
\end{equation}

So far we have found the modes describing quantum fluctuations outside the 
light cone from N
\begin{equation}
\phi_{-3lm}={\dot\sigma_0(\tau)\over (R_0\mu)^{1/2}} Y_{-3lm}(x^i).
\label{resultout}
\end{equation}
The modes $Y_{-3lm}$ are those of a scalar field of tachyonic 
$(mass)^2=k^2=-3$ \cite{jaumealex}
living in a (2+1) dimensional de Sitter 
space (\ref{desitter}).
If we want to preserve the O(3,1) symmetry of the 
bubble solution then we need to choose the Bunch-Davies vacuum
in the lower dimensional time-like $\tau=const$ sections. The
corresponding normalized modes are \cite{jaumealex}
\begin{equation}
Y_{-3lm}=-\left({\pi\Gamma(l-1)\over 4 \Gamma(l+3)}\right)^{1/2}
{1\over \cosh\rho}R^2_l(\tanh\rho) Y_{lm}(\theta,\varphi),
\label{dem}
\end{equation}
where $Y_{lm}$ are the spherical harmonics and $R^{\lambda}_{\nu}
(x)\equiv
{\rm P}^{\lambda}_{\nu}(x)-(2i/\pi){\rm Q}^{\lambda}_{\nu}(x)$. Here P and 
Q are the Legendre functions on the cut $-1<x<1$.

In order to assess the effect of these fluctuations in our Universe today, 
we must first analytically continue to the interior of the light-cone,
\begin{equation}
\phi_{-3lm}=
{\dot\sigma(t)\over (R_0\mu)^{1/2}}Y_{-3lm}(r,\theta,\varphi)\label{demol}
\end{equation}
where now the analytically continued harmonics can be cast, after some
algebra \cite{gr}, into the form
\begin{equation}
Y_{-3lm}=\left({\Gamma(3+l)\Gamma(l-1)\over 2}\right)^{1/2}
{P^{-l-1/2}_{3/2}(\cosh r)\over \sqrt{\sinh r}}Y_{lm}(\theta,\varphi).
\label{demo}
\end{equation}
Here we have used Eqns. 8.738.2 and 8.732.5 of Ref. \cite{gr}.
The Legendre functions can be given in terms of elementary functions.
For $l=0, 1$ and 2 they are
\begin{equation}
P^{-1/2}_{3/2}=(2\pi \sinh r)^{-1/2}\sinh(2r),
\label{legendre}
\end{equation}
$$
P^{-3/2}_{3/2}={1\over \Gamma(5/2)}\left({\sinh r\over 2}\right)^{3/2},
$$
$$
P^{-5/2}_{3/2}=\left({2\over\pi}\right)^{1/2}{1 \over 8(\sinh r)^{5/2}}
\left[{1\over 12}\sinh(4r)-{2\over 3}\sinh(2r)+ r\right],
$$ 
and for higher $l$ they can be obtained through the well known recurrence 
relations.

Several comments should be made. First of all, for $l=0$ and $l=1$ 
the normalization factor in (\ref{demo})
diverges. This is not a problem, since 
these modes do not contribute to observables. They simply 
correspond to spacetime translations of the 
nucleation event \cite{jaumealex}. 
Second, the modes are real on the open chart, and 
hence their Klein-Gordon norm vanishes there.
This is not a problem either, because the open 
hyperboloids are not good Cauchy surfaces \cite{sasaki}. 
Finally, the analytic continuation
of equation (\ref{living}) with $k^2=-3$ tells us that
$$
\Delta Y_{-3lm}=+3 Y_{-3lm},
$$
and so the eigenvalue of the Laplacian has the
`wrong' sign. Not surprisingly, 
the modes diverge exponentially for large  $r$. 
As we shall see, this divergence does not 
appear in the physical effect.

\section{From scalar to tensor modes}

The wall fluctuations induce scalar field fluctuations 
of the form (\ref{demol}) in the open FRW universe. These will locally 
advance or retard by an amount
\begin{equation}
\delta t={\phi_{-3lm}\over \dot\sigma}=(R_0\mu)^{-1/2}
Y_{-3lm}(r,\theta,\varphi)
\label{deltat}
\end{equation}
the time at which the universe reheats. Deformations of the reheating surface 
generically induce density fluctuations and perturbations 
in the microwave background. It turns out that the 
particular modes (\ref{demol}) do not cause density perturbations \cite{juan},
but, as we shall see, they do affect the microwave 
background just like gravitational waves do.

A nice framework to study deformations of the constant scalar field surfaces 
is that of Ref. \cite{ellis}. One defines a `fluid' velocity
\begin{equation}
u_{\mu}\equiv {\sigma,_{\mu}\over (-\sigma,_{\mu}\sigma,^{\mu})^{1/2}}
\label{velocity}
\end{equation}
orthogonal to the constant field surfaces, and projects its covariant 
derivative $u_{\mu;\nu}$ onto these surfaces:
$$
u_{\mu|\nu}\equiv (\delta_{\mu}^{\rho}+u_{\mu}u^{\rho})u_{\rho;\nu}.
$$
One can separate $u_{\mu|\nu}$ into a symmetric and an antisymmetric 
part. The antisymmetric part is called vorticity and
it can be shown that it vanishes for a four vector of the form 
(\ref{velocity}). As a consequence, the projected covariant derivative
$u_{\mu|\nu}$ coincides with the intrinsic covariant 
derivative on the surfaces $\sigma=const$ \cite{ellis}.
The symmetric tensor $K_{\mu\nu}=u_{\mu|\nu}$ is also known as the 
extrinsic curvature. Its trace $K_{\mu}^{\mu}\equiv\Theta$ 
is the expansion, and in the unperturbed FRW 
$\Theta=3(\dot a/a)$. 
A straightforward calculation shows that under perturbations
of the form (\ref{demol}) the expansion does not change \cite{jaumealex}.
The traceless part of $K_{\mu\nu}$
is the shear tensor 
$\sigma_{\mu\nu}=K_{\mu\nu}-(\Theta/3)(g_{\mu\nu}+u_{\mu}u_{\nu})$. 
To linear order in perturbations and in the coordinate 
system (\ref{interior}), we have
\begin{equation}
u_{\mu}=(-1,Y,_{i}),
\end{equation}
where $Y$ stands for $Y_{klm}(x^i)$, with $x^i=(r,\theta,\varphi)$.
Then $\sigma_{00}=\sigma_{0i}=0$ and 
\begin{equation}
\sigma_{ij}=Y_{|ij}-Y\gamma_{ij}.
\label{shear}
\end{equation}
Here $\gamma_{ij}$ is the metric on the unit spacelike hyerboloid (\ref{open}).
In addition to being traceless, the shear tensor for $k^2=-3$
is transverse $\sigma_{ij|}\ ^j=0$, just like a tensor mode \cite{sasaki}.

This immediately suggests going to a new coordinate system which straightens 
out the constant scalar field surfaces, while still remaining in a synchronous 
gauge
$$
t'=t+\delta t,\quad x^{i'}=x^i-\gamma^{ij}{\dot a\over a}\delta t_{|j}.
$$
Here $\delta t$ is given by (\ref{deltat}). In this new gauge 
$\sigma=\sigma(t')$ is constant on $t'=const.$ surfaces, but the metric 
reads
$$
ds^2=-dt^2+(\gamma_{ij}+h_{ij})dx^idx^j.
$$
Here
\begin{equation}
h_{ij}=-2E\sigma_{ij},\label{hij}
\end{equation}
and, during inflation
\begin{equation}
E={\dot a \over a}(R_0\mu)^{-1/2}.\label{e}
\end{equation}
Once in the new transverse and traceless gauge, 
the matter distribution looks perfectly smooth.
It is now legitimate to
evolve $h_{ij}$ with the usual equation for tensor perturbations 
through the entire cosmic evolution \cite{slava}
\begin{equation}
\ddot h_{ij} +3{\dot a\over a} \dot h_{ij}- 
{1\over a^2}(\Delta h_{ij} + 2h_{ij})=0.
\label{slava}
\end{equation}
Note that although $\Delta Y=+3Y$, the 
corresponding tensor mode, derived from (\ref{hij}) and (\ref{shear})
satisfies
$$
\Delta h_{ij}=- 3 h_{ij},
$$ 
with the `correct' sign for the Laplacian eigenvalue.
For $l=0$ and $l=1$, it can be readily checked that $\sigma_{ij}=0$,
and so the tensor mode only exists for $l>1$, as expected of 
gravitational waves \cite{allen}.

Introducing (\ref{hij}) in (\ref{slava}) we can calculate the 
evolution of the amplitude $E$ throughout the different stages of expansion. 
In terms of conformal time ($d\eta=a^{-1} dt$), we have
$$
E''+2{a'\over a}E+E=0.
$$
During inflation $E$ is given by (\ref{e}) and it tends to a constant,
$$
E={H\over \sqrt{R_0\mu}},
$$
where $H=(\Lambda/3)^{1/2}$.
In the radiation era $\Omega=1$ to very high acuracy, and it can be 
checked that as a result $E$ stays constant. 

During the matter era
$a\sim t^{2/3}\sim \eta^2$, so
$$
E''+{4\over\eta}  E'+E=0.
$$
This can be solved in terms of Bessel functions. For small $\eta$
$$
E={H\over\sqrt{R_0\mu}}(1-{\eta^2\over 8}+...).
$$
From (\ref{friedman}) we have  $\eta=2(1-\Omega)^{1/2}$, and
in what follows we shall concentrate in the case $(1-\Omega)<<1$.
(The general case can be treated numerically along the lines of 
Refs. \cite{lyth,yama}).

The amplitude of microwave fluctuations due to $h_{ij}$ is given by the 
well known Sachs-Wolfe formula
$$
{\delta T\over T}={1\over 2}\int_0^{r_{ls}} {dh_{rr}\over d\eta} dr,
$$
where $r_{ls}\approx 2(1-\Omega)^{1/2}$ is the comoving distance to the 
surface of last scattering and $h'_{rr}$ is evaluated at $\eta=r_{ls}-r$.
Since $r_{ls}$ is assumed to be small, we can use the asymptotic form of the 
Legendre functions in ({\ref{demo}) to obtain
$$
Y_{-3lm}\approx \Gamma_l 2^{-(l+1)} r^l Y_{lm},
$$
where $\Gamma_l=[\Gamma(3+l)\Gamma(l-1)]^{1/2}/\Gamma(l+3/2)$
Using this form in (\ref{shear}) and (\ref{hij}) one immediately obtains
\begin{equation}
{\delta T \over T} \approx {H\over \sqrt{R_0\mu}}{\Gamma_l\over 8}
(1-\Omega)^{l/2}, \label{deltatemp}
\end{equation}
which for low spatial curvature is dominated by the quadrupole $l=2$.

It should be noted that the tensor mode $h_{ij}$ is pure gauge during 
inflation (as it should since we are neglecting the bubble's backreaction.)
However, the overall configuration taking the scalar field into account is not.
The $\sigma=const.$ surfaces have non-vanishing shear. As we evolve
$h_{ij}$ past the reheating surface, it gradually
ceases to be pure gauge, as can be
checked by expressing the evolved mode in the 
longitudinal gauge \cite{slava}.  
Also, it can be checked that although the scalar modes $Y_{-3lm}$ diverge
exponentially at large distances, the corresponding tensor components
$h_{rr}$ entering the Sachs-Wolfe formula do not.

\section{Conclusions}

We have calculated the amplitude of small fluctuations in the shape of 
bubbles nucleated during inflation. In the case of thin bubbles, 
the result takes the  simple form (\ref{demol}). Since the de Sitter modes
$Y_{-3lm}$ grow like the intrinsic radius of the bubble 
$r=R_0 \cosh \rho$ ($\rho R_0$ is the proper time coordinate on the bubble 
wall), 
the relative amplitude of
proper local radial wall displacements is given by
$$
{\delta r\over r}\sim (R_0^3\mu)^{-1/2},
$$
where $R_0$ is the radius of the bubble at the time of nucleation, given by
(\ref{r0}), and $\mu$ is the wall tension. This coincides by order of 
magnitude with the estimate of Ref.\cite{linde} in the case when the size 
of the bubbles at the time of nucleation is much smaller than the de Sitter
horizon $H^{-1}$.

The propagation of wall perturbations to the interior of the light cone 
gives rise to shear deformations of the 
reheating surface, and consequently, of the surface of last scattering.
The simplest way to study the cosmological evolution of such perturbations 
is to use a synchronous coordinate system in which matter looks 
perfectly smooth. Somewhat surprisingly, the metric fluctuations in
this gauge are transverse and traceless, just like usual gravitational
waves. This transmutation of scalar into tensor-like modes is only 
possible because of the peculiar (supercurvature)
eigenvalue of the Laplacian for the scalar
modes corresponding to wall fluctuations, $k^2=-3$ \cite{sasaki}.

For
$(\Omega-1)<<1$, the anisotropies of the microwave sky produced
by these waves 
are given in (\ref{deltatemp}). The dominant effect is in 
the quadrupole, with
$$
{\delta T\over T}\sim {H\over\sqrt{R_0\mu}}(1-\Omega).
$$
Within the limits of validity of our approximation 
[see (\ref{weak})], and unless $\Omega$ is too close to 1,
this can be much larger than the distortion produced by usual gravitational
waves produced during inflation, which is of order $G^{1/2}H$. 
The case with strongly gravitating domain walls may yield a different
result \cite{juan}, and is currently under investigation.

Finally, we would like to compare our results with those of Ref. \cite{juan}.
There it is claimed that the $k^2=-3$ modes produce no distortions of the
microwave background. Here we have calculated
the amplitude of such perturbations and their nonvanishing effect. 
In Ref. \cite{juan} the amplitude of a $k=0$ 
homogeneous fluctuation is estimated by considering the small change 
$\Delta S$ in
the instanton action when the radius of the bubble is changed by 
an amount $\delta a$.
This is used to estimate the `typical deviation' $\delta a$
as the one that corresponds to  
$\Delta S \sim 1$. However, the physical meaning of that prescription is
unclear, because the radius of the bubble cannot have a homogeneous 
fluctuation. In flat space, this would violate energy 
conservation, and a similar argument can be applied in curved space.
The only homogeneous radial fluctuations allowed 
in the thin wall limit are time translations of
the nucleation point.

\section*{Acknowledgements}

It is a pleasure to thank Alex Vilenkin, Takahiro Tanaka and Juan 
Garcia-Bellido for stimulatind discussions, and the Tufts Cosmology
institute for warm hospitality during the preparation of this work.
This work has been partially supported by CICYT under project
AEN95-0882, and by a NATO collaborative research grant.

\section*{Figure captions}
\begin{itemize}

\item{Fig. 1}  A scalar field potential $V(\sigma)$ which leads to 
open inflation. The universe is initially in the false vacuum phase
$\sigma_f$ when a bubble of $\sigma_t$ nucleates. Then the scalar
field slowly rolls down the hill until it reaches $\sigma_{rh}$,
at which point the universe reheats. We separate the
potential into a large part $\Lambda/8\pi G$ and a small part with
the barrier feature, since we want to neglect the self 
gravity of the bubble. The top of the barrier is denoted by $\sigma_m$.

\item{Fig. 2}  A conformal diagram of spacetime in the presence of a bubble.
The nucleation event is marked as N.
Region I corresponds to an open FRW which inflates and eventually becomes
our observable universe. 
The trajectory of the domain wall is marked 
as $w$. It lies in Region II, which is covered by the chart
(\ref{exterior}). The spacelike hypersurface $\rho=0$, connecting
N with the antipodal point A,
is a good cauchy surface for the entire 
spacetime. Clearly, no such surfaces exist in Region I. Region III,
the interior of the light-cone from A, is uninteresting for our purposes.  

\item{Fig. 3} The effective potential $U_{eff}$ as a function of 
conformal radius $\eta$. The narrow well at $\eta_w$ corresponds 
to the location of the bubble wall, where the effective mass
$m^2(\sigma)$ is negative.

\end{itemize}

\newpage
\onefig{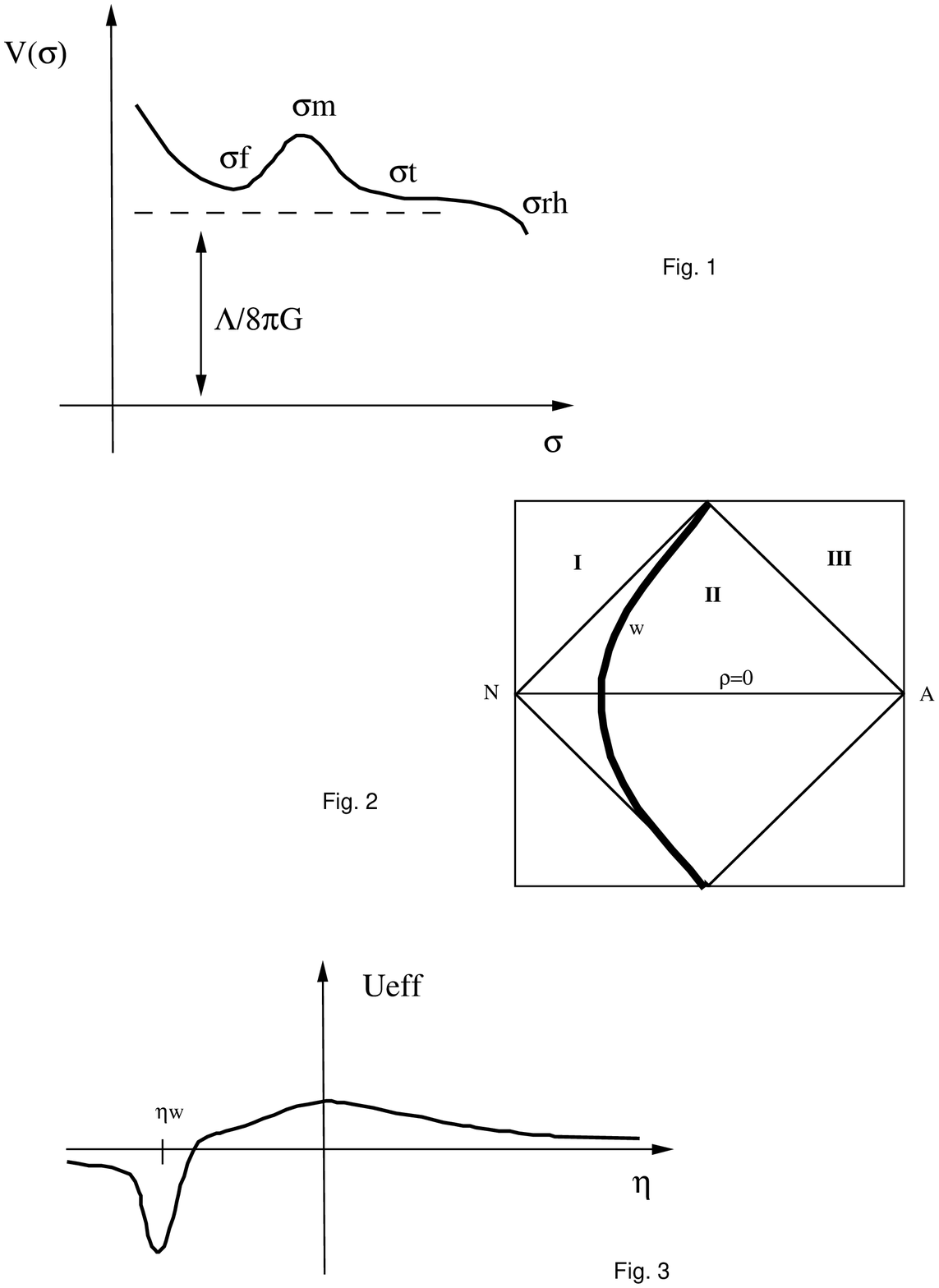}{}

\end{document}